\documentclass[pdflatex,sn-mathphys-num]{sn-jnl}

\usepackage{graphicx}%
\usepackage{multirow}%
\usepackage{amsmath,amssymb,amsfonts}%
\usepackage{amsthm}%
\usepackage{mathrsfs}%
\usepackage[title]{appendix}%
\usepackage{xcolor}%
\usepackage{textcomp}%
\usepackage{manyfoot}%
\usepackage{booktabs}%
\usepackage{algorithm}%
\usepackage{algorithmicx}%
\usepackage{algpseudocode}%
\usepackage{listings}%
\usepackage{lineno}
\usepackage{ulem}

\theoremstyle{thmstyleone}%
\theoremstyle{thmstyletwo}%

\theoremstyle{thmstylethree}%

\begin{document}

\title[Article Title]{Prevention of Yb adsorption by paraffin coating}

\author*[1,4]{\fnm{Kanta} \sur{Asakawa}}\email{asakawa-kanta-hn@ynu.ac.jp}

\equalcont{These authors contributed equally to this work.}

\author[1]{\fnm{Taichi} \sur{Kobayashi}}
\equalcont{These authors contributed equally to this work.}

\author[2]{\fnm{Masami} \sur{Yasuda}}
\author[3]{\fnm{Takao} \sur{Aoki}}
\author[1]{\fnm{Atsushi} \sur{Hatakeyama}}

\affil*[1]{\orgdiv{Department of Applied Physics}, \orgname{Tokyo University of Agriculture and Technology}, \orgaddress{\city{Koganei}, \state{Tokyo}, \postcode{184-8588}, \country{Japan}}}

\affil[2]{\orgdiv{National Metrology Institute of Japan (NMIJ)}, \orgname{National Institute of Advanced Industrial Science and Technology (AIST)}, \orgaddress{ \city{Tsukuba}, \state{Ibaraki}, \postcode{ 305-8563}, \country{Japan}}}

\affil[3]{\orgdiv{Department of Applied Physics}, \orgname{Waseda University}, \orgaddress{\street{Okubo}, \city{Shinjuku, Tokyo}, \postcode{169-8555}, \country{Japan}}}
\affil[4]{\orgdiv{Department of Physics, Graduate School of Engineering Science}, \orgname{Yokohama National University}, \orgaddress{\street{Tokiwadai}, \city{	Hodogaya-ku, Yokohama}, \postcode{ 240-8501}, \country{Japan}}}

\abstract{Ytterbium (Yb) is used in cold-atom systems,  including magneto-optical traps and optical lattice clocks. However, the long-term operation of such systems may be associated with substantial degradation of optical transmittance through vacuum chamber viewports due to Yb adsorption. Here, we show that coating the surface with tetracontane effectively suppresses such adsorption.}

\maketitle
 
\section{Introduction}\label{sec1}
Atomic  ytterbium (Yb) beams have many applications in laser cooling and trapping, including magneto-optical trapping (MOT) and optical lattice clocks. Such beams are typically generated by ovens operating at about 700 K \cite{Kawasaki_2015,letellier2023loading}. To ensure \textcolor{black}{an} efficient  MOT of Yb atoms, the atomic beam is often decelerated using a Zeeman slower, which involves introducing a counter-propagating laser beam against the Yb atomic beam. However, Yb atoms tend to accumulate on the viewport used to introduce the decelerating laser, reducing the  optical transmittance \cite{Kawasaki_2015,Xu_2016,10.1063/5.0140774}.   Yb atoms adsorbed onto the viewport are not readily removed by heating. In an effort to avoid this issue, a window heated to around 600 K is often placed inside the vacuum chamber in front of the viewport\cite{Kawasaki_2015}. However, this adds to the complexity of the apparatus. Additionally, radiation from the heated window degrades the performance of optical lattice clocks \cite{heo2022evaluation}.Future clocks must be both small and transportable. A heated window increases the space and power requirements, making miniaturization difficult \cite{yasudaLaser}.

 An alternative approach to prevent Yb deposition on the viewport is to apply a coating of a material that is resistant to Yb adsorption.  Here, we tested a paraffin coating (C$_{\rm{n}}$H$_{\rm{2n+2}}$). Paraffin is widely used to coat the walls of alkali-metal vapor cells. Such coatings suppress interactions between alkali-metal atomic vapors and the walls themselves \cite{studyofwallcoatings}, and reduce adsorption and spin relaxation of alkali-metal atoms that collide with the walls \cite{bouchiat,asakawaPRA}. We used X-ray photoelectron spectroscopy (XPS) to explore whether coating the viewport with tetracontane (C$_{40}$H$_{82}$), a representative paraffin, suppresses Yb adsorption from an atomic Yb beam.

\section{Experimental}\label{sec2}
\textcolor{black}{The paraffin used in this study was tetracontane because it is relatively readily available, has a vapor pressure low enough for use in UHV, and is known to be effective as an anti-adsorption and anti-spin-relaxation coating material\cite{ulanski,sekiguchi2018,asakawaPRA}.} \textcolor{black}{To avoid charging up, which reduces the accuracy of XPS measurements, Si was used as the substrate, which is more conductive than glass substrates. Note that the performance of the tetracontane coating is almost independent of the substrate material \cite{sekiguchi2018}}. A tetracontane thin film with a thickness of 300 nm was deposited on an Si(100) substrate. 
\textcolor{black}{The transmittance of a fused-quartz substrate with a 450-nm-thick tetracontane film, fabricated in advance, was measured to be 43\% for the laser with a wavelength of 399 nm, the wavelength used for Zeeman slowers. Therefore, a thickness of 300 nm examined in this experiment can provide adequate transmittance for use as a window for introducing laser. Because the paraffin coating retains its performance at thicknesses down to 100 nm\cite{atutov2018paraffin}, we believe that thinner (and thus more transparent) coatings can be used without degrading the anti-adsorption performance.}The deposition was conducted in a high-vacuum chamber with a base pressure of $\sim 2\times 10^{-4}$ Pa at a deposition rate of  0.35 nm/s by evaporating tetracontane from a crucible at 200 $^\circ$C, while maintaining the sample at room temperature. \textcolor{black}{During deposition, the film thickness was monitored with a thickness monitor, which was calibrated by using an atomic force microscope.} The sample was then transferred to the XPS measurement chamber through ambient air. \textcolor{black}{
Because tetracontane remains relatively clean even after air exposure\cite{sekiguchi2018,asakawaPRA}, we did not perform surface treatments to remove contaminants such as annealing, as such treatment may cause tetracontane to segregate. Also, it is practical to use air-exposed coatings in cold-atom systems because loading a coated window without air-exposure requires an in-situ deposition system with a transfer system, which increases the complexity of the apparatus significantly.} The base pressure of the XPS measurement chamber was lower than $2\times 10^{-7}$ Pa. \textcolor{black}{The pressure in the chamber did not change by more than $1\times 10^{-8}$ Pa upon loading the sample, which corresponds to the resolution of the pressure gauge. This confirms that degassing from the film was negligible.}
\begin{figure}
	\centering
	\includegraphics[width=0.7\linewidth]{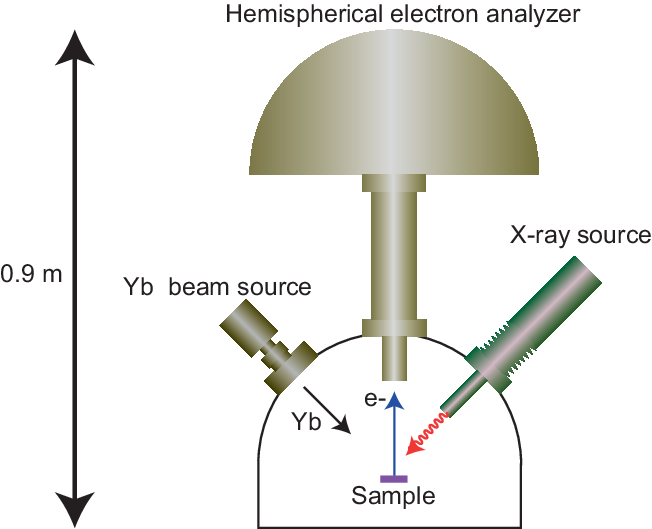}
	\caption{Schematic illustration of the XPS apparatus.}
	\label{fig:xps}
\end{figure} 

Figure \ref{fig:xps} shows a schematic of the XPS measurement chamber, which featured an Al K$\alpha$ X-ray source, a hemispherical electron analyzer, and an atomic Yb beam. The XPS measurement was conducted by irradiating the  sample with Al K$\alpha$ X-ray (1486.6 eV) and energy-analyzing the emitted photoelectrons using the hemispherical electron analyzer. The resulting spectra provided information on the atomic composition, depth distribution, and chemical states of the near-surface region of the sample. All XPS spectra were acquired in a normal-emission geometry, in which the surface-normal direction of the sample was aligned with the axis of the electron analyzer. The atomic Yb beam source featured a Yb reservoir and an array of 45 capillaries, each with an inner diameter of 0.3 mm, outer diameter of 0.4 mm, and length of 10 mm, through which Yb atoms were introduced into the UHV chamber to form a collimated beam.  The reservoir and capillaries were heated to 450 and 550 $^\circ$C, respectively. \textcolor{black}{These temperatures were measured with thermocouples placed near the heaters.} Under these conditions, the beam flux intensity and areal flux density at the sample position were designed to be \textcolor{black}{$\leq 6\times 10^{14}$} atoms/s and \textcolor{black}{$\leq5\times10^{14}$}  atoms/cm$^2\cdot$s.  \textcolor{black}{ It should be noted that the reservoir temperature was measured in the vicinity of the heater (probably the hottest point), and the temperature may vary by several tens of degrees depending on the measurement position. Since the beam flux intensity is primarily determined by the temperature of the coldest point in the reservoir,  the actual beam flux  may be significantly smaller  than these values.}

\section{Results and Discussion}\label{sec3}

\begin{figure}
	\centering
	\includegraphics[width=0.9\linewidth]{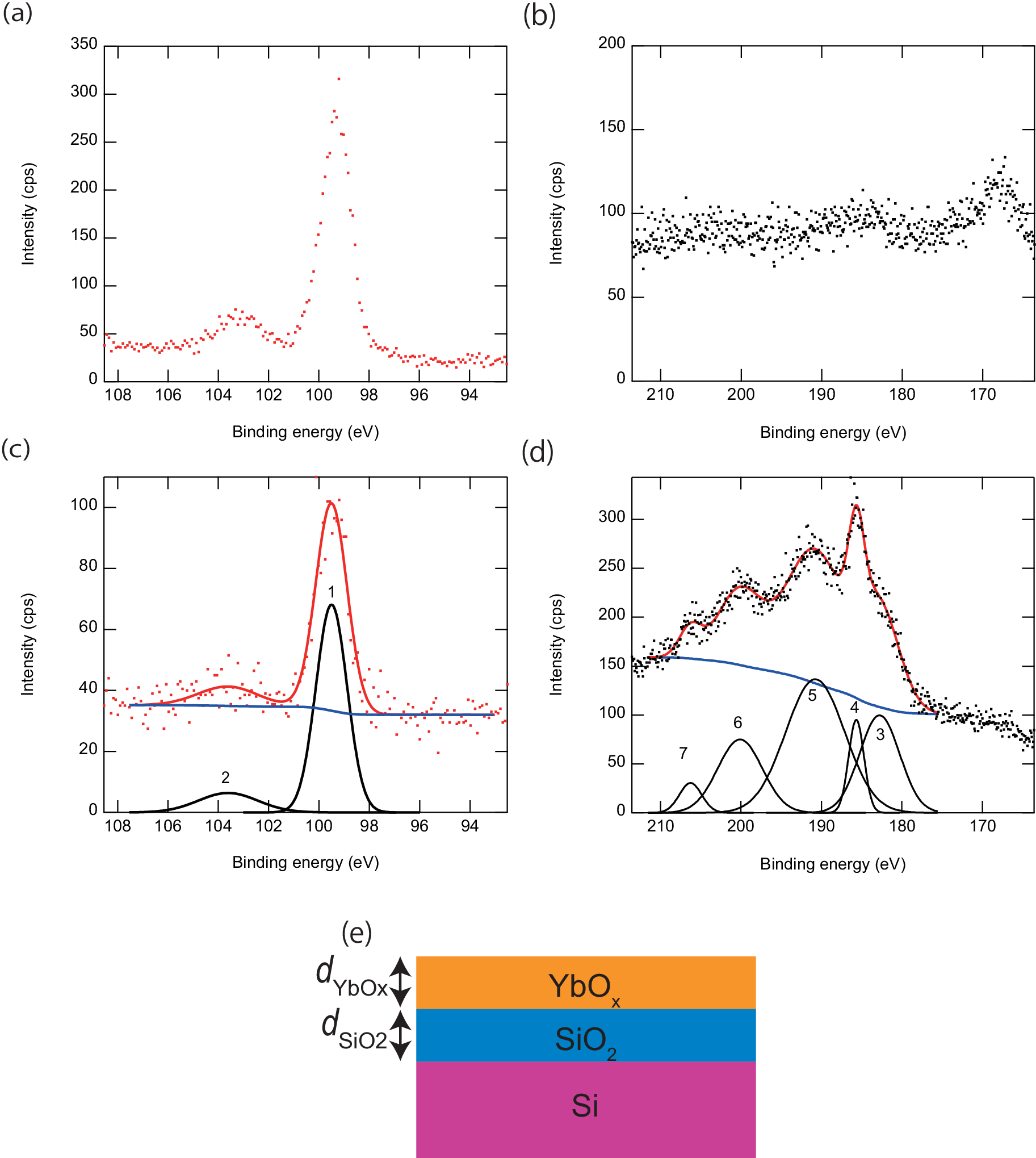}
	\caption{ \textcolor{black}{The XPS spectra of (a) Si 2p and (b) Yb 4d regions of the  native oxide layer of Si(100)},  (c) Si 2p and (d) Yb 4d regions obtained after  Yb atomic beam irradiation of the native oxide layer of Si(100) for 80 minutes, and (e) a suggested model of the surface layer structure.}
	\label{fig:ybonsio2}
\end{figure}

To estimate the adsorption rate of Yb atoms on a SiO$_2$ surface -- a typical material used for vacuum chamber viewports -- we irradiated a Yb atomic beam onto a Si(100) substrate covered with a native SiO$_2$ layer at room temperature. \textcolor{black}{Figures \ref{fig:ybonsio2}(a) and (b) show the XPS narrow scan spectra of the Si 2p and Yb 4d regions, respectively, before Yb beam irradiation. The binding energy scale was calibrated using the Si 2p peak at 99.5 eV.  The small peak at 167 eV in Figure \ref{fig:ybonsio2}(b) is a satellite peak of the Si 2s peak at 151 eV. No peak was observed at the position where the Yb 4d peaks should appear (175-210 eV). } Figure \ref{fig:ybonsio2}(c) and (d) show the XPS narrow scan spectra of the Si 2p and Yb 4d regions, respectively,  after 80 min of Yb beam exposure.   In the XPS spectra, the low binding energy sides of the peaks are superimposed on background signals attributable to inelastic scattering; these were subtracted using the Shirley method. Each background-subtracted  spectrum was fitted using two Gaussian functions. The peaks labeled as "Peak 1" and "Peak 2"  in Fig. \ref{fig:ybonsio2}(c) correspond to elemental Si and SiO$_2$, respectively. The fitted peak positions and integrated areas are summarized in Table \ref{tableSi}. The thickness $d_{\rm SiO_2}$ of the native SiO$_2$ layer was estimated using  the following equation:
 
 \begin{equation}
\frac{I_2}{I_1}=\frac{D_{\rm SiO_2}\int_{0}^{d_{\rm SiO_2}}\exp(\frac{-x}{l_{\rm Si2p,i}})dx}{D_{\rm Si}\exp(-\frac{d_{\rm SiO_2}}{l_{\rm Si2p,i}})\int_{0}^{\infty}\exp(\frac{-x}{l_{\rm Si2p,e}})dx}, \label{eq:dsio2}
\end{equation}
\begin{table}[htbp]
	\caption{Parameters of the Si 2p peaks.}
	\label{table:si2p}
	\centering
	\begin{tabular}{c|ccc}
		\hline\hline
		Peak Nos. & 1 & 2 \\
		\hline
		peak position (eV) & 99.5 & 103.6 \\
		peak area (cps$\cdot$eV) & $105.8\pm 3.1 $& $18.8\pm 4.3$ \\
		assignment & Si & SiO$_2$ \label{tableSi}\\
		\hline\hline
	\end{tabular}
\end{table}
where $I_1$ and $I_2$ are the areas of the peaks assigned to the Si and SiO$_2$ from Table \ref{tableSi}, respectively, $d_{\rm SiO_2}$ is the thickness of the SiO$_2$ layer, and $D_{\rm SiO_2}$ and $D_{\rm Si}$ are the densities of Si atoms in SiO$_2$ and Si crystals, respectively, which were $2.13\times10^{22}$ cm$^{-3}$\cite{HAUSER198159} and $5.00\times 10^{22}$ cm$^{-3}$\cite{Villars2023:sm_isp_sd_1622696}, respectively. $l_{\rm Si2p,e}$ and $l_{\rm Si2p,i}$ are the inelastic mean free paths of photoelectrons from the Si 2p orbital that were excited by the Al K$\alpha$ X-ray in Si and SiO$_2$, respectively. These values were estimated using the following approximation:\cite{seah}

\begin{equation}
	l_{\rm L,M}=A_{\rm M} E_{\rm L}^2+B_{\rm M} E_{\rm L}^{1/2}, \label{IMFP}
\end{equation}                                                                      where $l_{\rm L,M}$ is the inelastic mean free path of photoelectrons (in nm) and $E_{\rm L}$ is the kinetic energy (in eV). "L" refers to a photoelectron line, such as that of Si 2p. "M" is a material parameter: ${\rm M=e}$ for elements, ${\rm M=i}$ for inorganic compounds and ${\rm M=o}$ for organic compounds. Here, ${\rm M=e}$ corresponds to Si, ${\rm M=i}$ to SiO$_2$ and Yb oxide, and ${\rm M=o}$ to tetracontane. 
 The $A_{\rm M}$ and $B_{\rm M}$ values are listed in Table \ref{table:A,Bvalues}.  Substitution of $E_{\rm Si2p}=1387$ eV into Eq. (\ref{IMFP}) yielded calculated inelastic mean free paths of Si 2p photoelectrons in the Si and SiO$_2$ layers as $l_{\rm Si2p,e}=2.01$ nm and $l_{\rm Si2p,i}=3.58$ nm, respectively. By numerically solving Eq. (\ref{eq:dsio2}), we obtained $d_{\rm SiO_2}=0.75 $ nm.

\begin{table}[htbp]
	\caption{The $A_{\rm M}$ and $B_{\rm M}$\cite{seah} values.}

	\centering
	\begin{tabular}{ccc}
		\hline\hline
	M	 & $A_{\rm M}$ (nm$\cdot$eV$^{-2}$)& $B_{\rm M}$ (nm$\cdot$eV$^{-\frac{1}{2}}$)\\
		\hline
		e & 143 & 0.054 \\
	i & 641& 0.096 \\
	o & 31 & 0.087 \label{table:A,Bvalues}\\
		\hline\hline
	\end{tabular}
\end{table}

\begin{table}[htbp]
	\caption{The parameters of the Yb 4d peaks.}
	\label{table:yb4d}
	\centering
	\begin{tabular}{c|ccccc}
		\hline\hline
		Peak Nos. & 3 & 4 & 5 & 6 & 7 \\
		\hline
		peak position (eV) & 182.8 & 185.7 & 190.8 & 200.1 & 206.3 \\
		peak area (cps$\cdot$eV) & $622.3\pm81.8$ & $227.7\pm36.8$ & $1252.8\pm97.4$ & $519.9\pm55.5$ & $105.9\pm18.4$ \\
		assignment & metallic Yb & Yb$^{3+}$ &\multicolumn{3}{c}{ Yb 4d$_{3/2}$  and plasmon loss peaks}\label{tableYb} \\
		\hline\hline
	\end{tabular}
\end{table}
Peaks 3 and 4 in Figure \ref{fig:ybonsio2}(d) correspond to the Yb 4d$_{5/2}$ signals from metallic Yb and Yb$^{3+}$, respectively \cite{chastain1992handbook}. Peaks 5, 6, and 7 include contributions from both Yb 4d$_{3/2}$  and the plasmon loss peaks \cite{quantitative}. The binding energies and integrated areas of these peaks are summarized in Table \ref{tableYb}. The coexistence of neutral Yb and Yb$^{3+}$ indicates that the adsorbed Yb layer is partially oxidized. We therefore assume that a mixture of metallic Yb and Yb$_2$O$_3$, which we term YbO$_x$, covered the SiO$_2$ layer [Fig.~\ref{fig:ybonsio2}(e)].  Because the amounts of neutral Yb and Yb$^{3+}$ are proportional to $I_3$ and $I_4$, respectively (where $I_3$ and $I_4$ are the integrated areas of the Yb 4d$_{5/2}$ peaks of metallic Yb and Yb$^{3+}$), the volumes occupied by metallic Yb and Yb$_2$O$_3$ are proportional to $\frac{I_3}{D_{\rm Yb}}$ and $\frac{I_4}{D_{\rm Yb_2O_3}}$, respectively. Here,  $D_{\rm Yb}$ and $D_{\rm Yb_2O_3}$ are the density of Yb atoms in Yb and Yb$_2$O$_3$ crystals, which are estimated as  $D_{\rm Yb}=2.42\times 10^{22}$  cm$^{-3}$ \cite{BEAUDRY1974225} and $D_{\rm Yb_2O_3}=2.60\times 10^{22}$ cm$^{-3}$ \cite{osti_1202260}. Therefore, the density of Yb atoms in the adsorbed layer, $D_{{\rm YbO}_x}$, can be written as

\begin{equation}
D_{\rm YbO_x}=\frac{I_3+I_4}{\frac{I_3}{D_{\rm Yb}}+\frac{I_4}{D_{\rm Yb_2O_3}}},
\end{equation}
 Based on this analysis, we obtain: 
 
 \begin{equation}
 D_{{\rm YbO}_x}=2.48\times 10^{22}\  \rm{cm^{-3}}. \label{DYbOx}
 \end{equation}
 The thickness of the YbO$_x$ layer can be determined by comparing the experimentally obtained intensity ratio of the Si- and Yb-derived peaks with the calculated values based on assumed YbO$_x$ thicknesses, as expressed by the following equation:

\begin{eqnarray}
\frac{I_3+I_4}{I_1+I_2}&=&	\frac{S_{\rm Yb 4d_{5/2}}} {S_{\rm Si 2p}} \cdot
	\frac{f_{{\rm YbO}_{x}} }{
	 f_{\rm SiO_{2}}+f_{\rm Si}},\label{eq:Ybthickness}\\
		f_{{\rm YbO}_{x}}&=&	D_{{\rm YbO}_x}\cdot \int_0^{d_{{\rm YbO}_x}} \exp \left(-\frac{x }{ l_{\rm Yb4d,i}}\right)  dx,\nonumber\\
	f_{\rm SiO_{2}}&=&	D_{\rm SiO_2}\exp(-\frac{d_{{\rm YbO}_x}}{l_{\rm Si2p,i}}) \int_{0}^{d_{\rm SiO_2}} \exp\left(-\frac{x }{l_{\rm Si2p,i}} \right) dx,\nonumber\\
	f_{\rm Si}&=&	D_{\rm Si} \exp(-\frac{d_{{\rm YbO}_x}+d_{\rm SiO_2}}{l_{\rm Si2p,i}}) \int_{0}^{\infty} \exp \left(-\frac{x }{ l_{\rm Si2p,e}}\right)  dx,\nonumber
\end{eqnarray}                                                                                                                                                    
where $S_{\rm Yb 4d_{5/2}}$ and $S_{\rm Si 2p}$ are the normalized photoionization cross-sections of the Yb 4d$_{5/2}$ and Si 2p orbitals for Al K$\alpha$ X-ray, respectively, and $l_{\rm Yb4d,i}$ is the inelastic mean free path of Yb 4d photoelectrons in the YbO$_x$ layer. Here, $S_{\rm Yb 4d_{5/2}}=6.85$ and $S_{\rm Si 2p}=0.865$ \cite{SCOFIELD1976129}. Substitution of  $E_{\rm Yb4d}=1304$ eV into Eq.(\ref{IMFP}) yielded $l_{\rm Yb4d}=3.47 $ nm.  A numerical solution of Eq. \ref{eq:Ybthickness} revealed that the thickness $d_{{\rm YbO}_x}$ of the adsorbed layer was 3.2 nm.  From Eq. (\ref{DYbOx}), the areal density of Yb atoms was calculated to be $7.9\times 10^{15}$ cm$^{-2}$. 

\begin{figure}
	\centering
	\includegraphics[width=0.9\linewidth]{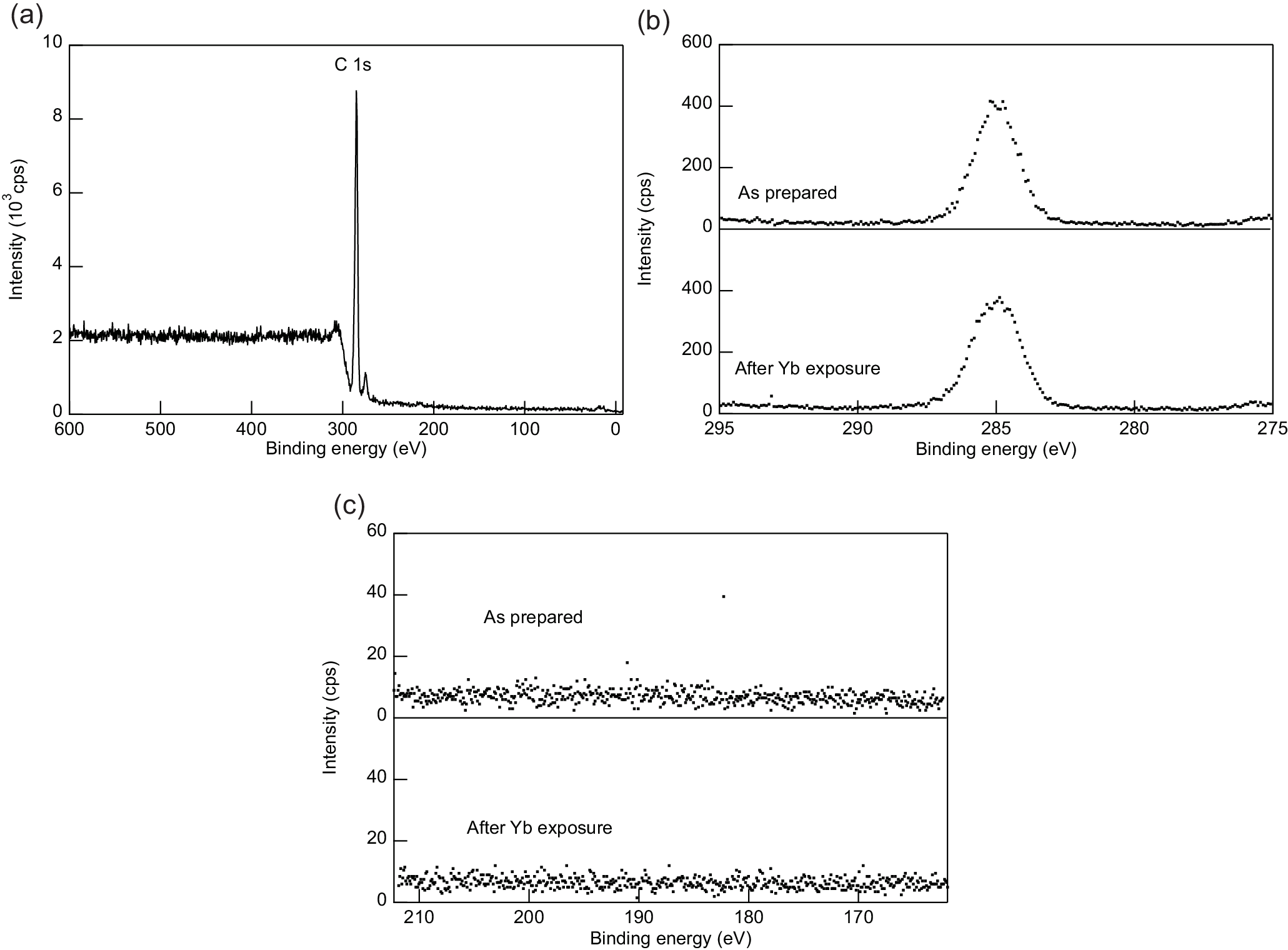}
	\caption{(a) Wide-scan XPS spectrum of the as-prepared tetracontane film. (b) Narrow scans of the C 1s peaks before and after Yb irradiation, and (c) narrow scans of the Yb 4d region before and after Yb irradiation.}
	\label{fig:xps-yb-on-tetracontane}
\end{figure}

Figure~\ref{fig:xps-yb-on-tetracontane}(a) shows the wide-scan XPS spectrum of the tetracontane film. The binding energy scale was calibrated using the C~1s peak at 285.0~eV. Because the thickness of the tetracontane film was substantially greater than the XPS probing depth, no Si- or O-derived peaks were observed. \textcolor{black}{The absence of O-derived peaks also means that there are no oxygen-containing contaminants on the surface, such as oxygen-containing organic molecules, CO, or H$_{2}$O.}
Figure~\ref{fig:xps-yb-on-tetracontane}(b) and \ref{fig:xps-yb-on-tetracontane}(c) present the narrow-scan spectra of the C~1s and Yb~4d regions, respectively, acquired before and after 80 min of irradiation with the Yb beam.  The C~1s peak area, \( I_{\rm C1s} \), listed in Table~\ref{table2}, did not significantly change after Yb irradiation. This indicates that the thickness of the adsorbed Yb layer was much smaller than the inelastic mean free path of C 1s photoelectrons in the adsorbed Yb layer. Because the Yb~4d peak was not distinct in the spectrum obtained after irradiation, its intensity was estimated by integrating the spectral region corresponding to the Yb~4d$_{5/2}$ peak (175--190~eV), rather than applying Gaussian fitting. The results are summarized in Table~\ref{table3}. The increase in integrated spectral intensity, \( \Delta I_{\rm Yb4d5/2} \), was \( 11.5 \pm 6.1 \)~cps$\cdot$eV.
 Eq. (\ref{IMFP}) revealed that the inelastic mean free path of the C~1s photoelectrons in tetracontane, \( l_{\rm C1s,o} \), was  3.0~nm. The density of carbon atoms in  the tetracontane film, \( D_{\rm tc} \), is \( 1.15 \times 10^{22}~\mathrm{cm}^{-3} \)~\cite{weber}. The areal density of C atoms within the probed depth is written as $ l_{\rm C1s,o}D_{\rm tc}$.
The experimentally obtained intensity ratio $\Delta I_{\rm Yb4d_{5/2}}/ I_{\rm C1s} $, yielded the estimated areal density of the Yb atoms, \( A_{\rm Yb} \), as given by the following expression:

\begin{equation}
	A_{\rm Yb}=l_{\rm C1s,o}D_{\rm tc}\cdot\frac{\Delta I_{\rm Yb4d_{5/2}}}{I_{\rm C1s} }\cdot\frac{S_{\rm C1s}}{S_{\rm Yb4d_{5/2}}} ,
\end{equation}
where $S_{\rm C1s}$ is the normalized photoionization cross-section for C 1s, which is 1.00 \cite{SCOFIELD1976129}. We assumed that the thickness of the adsorbed Yb layer was much smaller than \( l_{\rm C1s,o} \). Accordingly, we found that $	A_{\rm Yb}=(7.5\pm4.0)\times 10^{12}$ cm$^{-2}$, equivalent to  $ 0.09\pm0.05$ \% of the amount of Yb atoms  on the  surface of the Si native oxide surface. Note that, when estimating the Yb~4d$_{5/2}$ peak intensity, we neglected any contributions from the inelastic background and the Yb 4d$_{3/2}$ peak, which may lead to an overestimation of the adsorbed Yb amount. Therefore, our results indicate that Yb adsorption was suppressed to below 0.14\% by coating the SiO$_2$ surface with tetracontane.

\begin{table}[htbp]
	\caption{The  C1s peak areas before and after Yb irradiation.}
	
	\begin{tabular}{c|cc}
		
		\hline
		\hline
		&Before irradiation & After irradiation\\
		\hline

		$I_{\rm C1s}$ (cps$\cdot$eV)	& $771\pm 6$& $780\pm 5$ \label{table2} \\	\hline	\hline		
	\end{tabular}

\end{table}

\begin{table}[htbp]
	\caption{ The integrated spectral intensities $ I_{\rm Yb4d5/2}$ of the Yb 4d$_{5/2}$ region (175--190 eV) before and after Yb irradiation.}
	
	\begin{tabular}{c|cc}
		
		\hline
		\hline
		&Before irradiation & After irradiation\\
		\hline

 $I_{\rm Yb4d_{5/2}}$ (cps$\cdot$eV)	& $94.3\pm 4.2$& $105.7\pm 4.4$\label{table3} \\	\hline	\hline		
	\end{tabular}

\end{table}
\section{Conclusions}\label{sec4}

Suppression of Yb adsorption by tetracontane coating was investigated using an atomic Yb beam and XPS. The results showed that, by coating the SiO$_2$ surface with tetracontane film, the amount of Yb adsorption was suppressed to below 0.14\%. This finding will aid miniaturization and improve the power efficiency of cold-atom systems such as Yb optical lattice clocks. \textcolor{black}{For future applications in optical lattice clocks, it is necessary to investigate the stability of the tetracontane coating against long-term exposure to Yb, which is a future challenge. Another future challenge is to verify whether the prevention of adsorption by the tetracontane coatings can be applied to other atoms used in cold atom systems, such as Sr and Ca.}

During preparation of this manuscript, we became aware of a related work by Ajith \textit{et al.}\cite{AjithSurface}. In their study, they demonstrated that a polydimethylsiloxane coating -- a common wall-coating material -- exhibits a very low adsorption rate for Yb and Fe, which persists down to 200 K, consistent with the results presented here.
\section{Acknowledgements}
This research was supported by JST Moonshot R\&D (Grant Number JPMJMS2268).


\end{document}